\documentstyle[twoside,fleqn,espcrc2]{article}

\renewcommand{\mathrm}[1]{{\rm #1}}

\newcommand{\AmS}{{\protect\the\textfont2
  A\kern-.1667em\lower.5ex\hbox{M}\kern-.125emS}}

\def\Preprint{\vspace*{-7.0cm}   %\noindent hep-ph/0010XXX
  \noindent FTUV/00-1017 \\ %\mbox{}\hfill 
  IFIC/00-60 \\  %\mbox{}\hfill October 2000 \\ 
  \vspace{4.4cm}}

\def\NP{Nucl. Phys.}

\def\PL{Phys. Lett.}
\def\PRL{Phys. Rev. Lett.}
\def\PR{Phys. Rev.}

\newcommand{\eqn}[1]{(\ref{#1})}
\newcommand{\be}{\begin{equation}}
\newcommand{\ee}{\end{equation}}
\newcommand{\no}{\nonumber}

\newcommand{\ba}{\begin{array}{c}}
\newcommand{\bat}{\begin{array}{cc}}
\newcommand{\ea}{\end{array}}
\newcommand{\beqn}{\begin{eqnarray}}
\newcommand{\eeqn}{\end{eqnarray}}

\newcommand{\bi}{\begin{itemize}}
\newcommand{\ei}{\end{itemize}}

\def\eps{\varepsilon}

\newcommand{\lsim}{~{}_{\textstyle\sim}^{\textstyle <}~}

\newcommand{\cL}{{\cal L}}

\newcommand{\cA}{{\cal A}}

\begin{document}
% declarations for front matter

\title{Theoretical status of $\varepsilon'/\varepsilon$
 %\thanks{
 %Work supported in part by the ECC, TMR Network $EURODAPHNE$
 %(ERBFMX-CT98-0169), and by
 %DGESIC (Spain) under grant No. PB97-1261.  }
 \thanks{Invited talk at the Fourth
  Intern. Conference: {\it Hyperons, Charm and Beauty Hadrons}
  (Val\`encia, 27--30 June 2000)}
}

\author{A. Pich\address{Departament de F\'{\i}sica Te\`orica, 
         IFIC, Universitat de Val\`encia -- CSIC, \\ 
         %Edifici d'Instituts d'Investigaci\'o de Paterna,
         Apt. Correus 22085, E--46071 Val\`encia, Spain}}
         %Dr. Moliner 50, E--46100 Burjassot, Val\`encia, Spain}}

\begin{abstract}
\noindent
The Standard Model prediction for $\varepsilon'/\varepsilon$ is updated,
taking into account the most recent theoretical developments.
The final numerical value,
$\varepsilon'/\varepsilon = (17\pm 6)\times 10^{-4}$,
is in good agreement with present measurements.
\end{abstract}

% typeset front matter
\maketitle
\Preprint

\section{INTRODUCTION}
\label{sec:introduction}

The CP--violating ratio  $\varepsilon'/\varepsilon$  constitutes
a fundamental test for our understanding of flavour--changing
phenomena within the Standard Model framework.
The experimental status has been clarified by the recent
KTEV \cite{ktev:99} and NA48 \cite{na48:00} measurements.
The present world average \cite{na48:00}, 
\be\label{eq:exp}
{\rm Re} \left(\varepsilon'/\varepsilon\right) =
(19.3 \pm 2.4) \cdot 10^{-4} \, ,
\ee
provides clear evidence for a non-zero value and,
therefore, the existence of direct CP violation.

The theoretical prediction has been rather controversial since
different groups, using different models or approximations,
have obtained different results. Although there was no
universal agreement on the $\varepsilon'/\varepsilon$ value
predicted by the Standard Model, it has been often claimed
that it is too small, failing to reproduce the experimental
world average by at least a factor of two.
This claim has generated a very intense theoretical activity,
searching for new sources of CP violation beyond the Standard Model.

%%%%%%%%%%%%%%% TABLE %%%%%%%%%%%%%%%%%%%%%%%
\begin{table}[bth]
\centering
\caption{Recent predictions on $\varepsilon'/\varepsilon$.
G (gaussian) and S (scanning) refer to different error
analyses.}
\label{tab:pred}
\vspace{0.2cm}
\begin{tabular}{cccc}
\hline
$B_6^{(1/2)}$ &
$\varepsilon'/\varepsilon\times 10^4$ & & Ref.
\\ \hline
$1.0\pm 0.3$ & $\ba 9.2\,{}^{+6.8}_{-4.0} \\ 1.4\to 32.7 \ea $
& $\ba {\rm (G)} \\ {\rm (S)} \ea $ & \cite{munich}
\\
$1.0\pm 1.0$ & $\ba 8.1\,{}^{+10.3}_{-9.5}\\ -13\to 37 \ea $
 & $\ba {\rm (G)} \\ {\rm (S)} \ea $ & \cite{rome}
\\
$1.3\to 1.8$ & $\ba 22\pm 8 \\ 9\to 48 \ea $ &
$\ba {\rm (G)} \\ {\rm (S)} \ea $ & \cite{trieste}
\\
$1.50\to 1.62$ & $6.8\to 63.9$ & (S) & \cite{dortmund}
\\
$1.0$ & $-3.2\to 3.3$ & (S) & \cite{dubna}
\\
$2.5\pm 0.4$ & $34\pm 18$ && \cite{BP:00}
\\
$1.0\pm 0.4$ & $4\pm 5$ && \cite{NA:00}
\\
$1.5$ & $7\to 16$ & & \cite{taipei}
\\
$1.55$ & $17\pm 6$ && \cite{PP:00a,PP:00b,PPS:00}
\\ \hline
\end{tabular}
\end{table}
%
%%%%%%%%%%%%%%%%%%%%%%%%%%%%%%%%%%%%%%%%%%%%%%%%

It has been pointed out recently \cite{PP:00a} that the theoretical 
short--distance evaluations of $\eps'/\eps$ had overlooked
the important role of final--state interactions (FSI) in
$K\to\pi\pi$ decays.
Although it has been known for more than a decade that the
rescattering of the two final pions induces a large correction
to the isospin--zero decay amplitude, this effect was not
taken properly into account in the theoretical predictions.
From the measured $\pi$-$\pi$ phase shifts one can easily infer
\cite{PP:00a,PP:00b,PPS:00} that
FSI generate a strong enhancement of the $\eps'/\eps$ prediction,
%by roughly a factor of two ,
providing a good agreement with the experimental value.

The following sections present a brief overview of the most important
ingredients entering the Standard Model prediction of $\eps '/\eps$.

\section{THEORETICAL FRAMEWORK}
\label{sec:theory}

In terms of the $K\to\pi\pi$ isospin amplitudes,
$\cA_I = A_I \, e^{\delta_I}$ ($I=0,2$),
\be
{\varepsilon^\prime\over\varepsilon} =
\; e^{i\Phi}\; {\omega\over \sqrt{2}\vert\eps\vert}\;\left[
{\mbox{Im}A_2\over\mbox{Re} A_2} - {\mbox{Im}A_0\over \mbox{Re} A_0}
 \right] \, .
\ee
Owing to the well-known ``$\Delta I=1/2$ rule'', $\eps'/\eps$ is
suppressed by the ratio
$\omega = \mbox{Re} A_2/\mbox{Re} A_0 \approx 1/22$.
The strong S--wave rescattering of the two final pions generates a
large phase-shift difference between the two isospin amplitudes,
making the phases of $\eps'$ and $\eps$ nearly equal. Thus,
\be
\Phi \approx \delta_2-\delta_0+\frac{\pi}{4}\approx 0 \, .
\ee
The CP--conserving amplitudes $\mbox{Re} A_I$, their ratio
$\omega$ and $\eps$ are usually set to their experimentally
determined values. A theoretical calculation is then only needed
for the quantities $\mbox{Im} A_I$.

One starts above the electroweak scale where the flavour--changing
process, in terms of quarks, leptons and gauge bosons, can be analyzed
within the usual gauge--coupling perturbative expansion
in a rather straightforward way.
Since $M_W$ is much larger than the long--distance
hadronic scale $M_K$, there are large short--distance logarithmic
contributions which can be summed up using the
Operator Product Expansion (OPE) \cite{WI:69} and the renormalization
group. The proper way to proceed makes use of modern
Effective Field Theory (EFT) techniques \cite{EFT}.

The renormalization group is used to evolve down in energy 
from the electroweak scale, where the top quark and the $Z$ and 
$W^\pm$ bosons are integrated out. That means that one changes to a
different EFT where those heavy particles are no longer 
explicit degrees of freedom. The new Lagrangian contains a tower of
operators constructed with the light fields only, which scale as
powers of $1/M_W$. The information on the heavy fields is hidden in
their (Wilson) coefficients, which are fixed by
``matching'' the high-- and low--energy theories at the point $\mu=M_W$.
One follows the evolution further to lower energies, using the
EFT renormalization group equations, until a new particle
threshold is encountered. Then, the whole procedure of integrating the
new heavy scale and matching to another EFT starts again.

%%%%%%%%%%%%%%%%%%%%%%%%%%  FIGURE  %%%%%%%%%%%%%%%%%%%
\begin{figure}[tbh]    
\setlength{\unitlength}{0.65mm} \centering   %.75
\begin{picture}(115,113)
\put(0,0){\makebox(115,113){}}
\thicklines
\put(0,101){\makebox(20,13){\large Scale}}
\put(29,101){\makebox(36,13){\large Fields}}
\put(75,101){\makebox(40,13){\large Eff. Theory}}
\put(0,103){\line(1,0){115}} {\large
\put(0,70){\makebox(20,27){$M_W$}}
\put(29,70){\framebox(36,27){$\ba W, Z, \gamma, g \\
     \tau, \mu, e, \nu_i \\ t, b, c, s, d, u \ea $}}
\put(75,70){\makebox(40,27){\vbox{Standard \\ Model}}}

\put(0,35){\makebox(20,18){$\lsim m_c$}}
\put(29,35){\framebox(36,18){$\ba  \gamma, g  \; ;\; \mu ,  e, \nu_i  
             \\ s, d, u \ea $}} 
\put(75,35){\makebox(40,18){$\cL_{\mathrm{QCD}}^{(n_f=3)}$,  
             $\cL_{\mathrm{eff}}^{\Delta S=1,2}$}}

\put(0,0){\makebox(20,18){$M_K$}}
\put(29,0){\framebox(36,18){$\ba\gamma \; ;\; \mu , e, \nu_i  \\ 
            \pi, K,\eta  \ea $}} 
\put(75,0){\makebox(40,18){$\chi$PT}}
\linethickness{0.3mm}
\put(47,32){\vector(0,-1){11}}
\put(47,67){\vector(0,-1){11}}
\put(51,59.5){OPE} 
\put(51,24.5){$N_C\to\infty$}}
\end{picture}
\caption{Evolution from $M_W$ to $M_K$.
  \label{fig:eff_th}}
\end{figure}
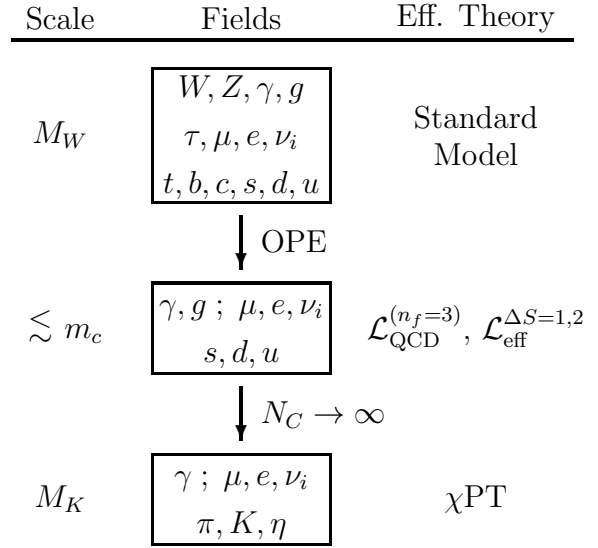
%%%%%%%%%%%%%%%%%%%%  END FIGURE %%%%%%%%%%%%

One proceeds down to scales $\mu < m_c$ and
gets finally an effective $\Delta S=1$ Lagrangian, defined in the
three--flavour theory \cite{GW:79,BURAS},
\be\label{eq:Leff}
 {\cal L}_{\mathrm eff}^{\Delta S=1}= - \frac{G_F}{\sqrt{2}}
 V_{ud}^{\phantom{*}}\,V^*_{us}\,  \sum_{i=1}^{10}
 C_i(\mu) \; Q_i (\mu) \; ,
 \label{eq:lag}
\ee
which is a sum of local four--fermion operators $Q_i$,
constructed with the light degrees of freedom, modulated
by Wilson coefficients $C_i(\mu)$ which are functions of the
heavy masses.
We have explicitly factored out the Fermi coupling $G_F$
and the Cabibbo--Kobayashi--Maskawa (CKM) matrix elements
$V_{ij}$ containing the usual Cabibbo suppression of $K$ decays.
The unitarity of the CKM matrix allows to write
%the Wilson coefficients in the form
%
\be
C_i(\mu) =  z_i(\mu) + \tau\ y_i(\mu) \; , 
\label{eq:Lqcoef} 
\ee
where 
$\tau = - V_{td}^{\phantom{*}} V_{ts}^{*}/V_{ud}^{\phantom{*}} V_{us}^{*}$.  
Only the $y_i$ components are needed to determine the CP--violating 
decay amplitudes.

The overall renormalization scale $\mu$ separates
the short-- ($M>\mu$) and long-- ($m<\mu$) distance contributions,
which are contained in $C_i(\mu)$ and $Q_i$, respectively.
The physical amplitudes are of course independent of $\mu$.
%; thus, the
%explicit scale (and scheme) dependence of the Wilson coefficients should
%cancel exactly with the corresponding dependence of the $Q_i$
%matrix elements between on-shell states.

Our knowledge of $\Delta S=1$ transitions has improved qualitatively
in recent years, thanks to the completion of the next-to-leading
logarithmic order calculation of the Wilson coefficients
\cite{buras1,ciuc1}.
All gluonic corrections of $O(\alpha_s^n t^n)$ and
$O(\alpha_s^{n+1} t^n)$ are known,
where $t\equiv\ln{(M_1/M_2)}$ refers to the logarithm of any ratio of
heavy mass scales $M_1,M_2\geq\mu$.
Moreover, the full $m_t/M_W$ dependence (at lowest
order in $\alpha_s$) is taken into account.

In order to predict physical amplitudes, however, one is still
confronted with the calculation of hadronic matrix elements of
the four--quark operators. This is a very difficult problem,
which so far remains unsolved.
Those matrix elements are usually parameterized
in terms of the so-called bag parameters $B_i$, which measure them
in units of their vacuum insertion approximation values.

To a very good approximation, the Standard Model prediction for
$\varepsilon'/\varepsilon$ can be written (up to global factors)
as \cite{munich}
\be
{\varepsilon'\over\varepsilon} \sim
\left [ B_6^{(1/2)}(1-\Omega_{IB}) - 0.4 \, B_8^{(3/2)}
 \right ]\, .
\label{EPSNUM}
\ee
Thus, only two operators are numerically relevant:
the QCD penguin operator $Q_6$ governs $\mbox{Im}A_0$
($\Delta I=1/2$), while $\mbox{Im}A_2$ ($\Delta I=3/2$)
is dominated by the electroweak penguin operator $Q_8$.
The parameter
\be
\Omega_{IB} = {1\over \omega}
{(\mbox{Im}A_2)_{IB}
\over \mbox{Im}A_0}
\label{eq:isospin}
\ee
takes into account isospin breaking corrections, which get enhanced
by the large factor $1/\omega$.

The value $\Omega_{IB}=0.25$ \cite{Omega,BG:87}
was usually adopted in all calculations.
Together with $B_i\sim 1$, this produces a large numerical cancellation
in eq.~\eqn{EPSNUM} leading to low values of $\eps'/\eps$
around $7\times 10^{-4}$.
The isospin--breaking correction coming from $\pi^0$-$\eta$  mixing
has been recently calculated at $O(p^4)$ in
Chiral Perturbation Theory ($\chi$PT), with the result
$\Omega_{IB}= 0.16\pm 0.03$ \cite{EMNP:00}.
This smaller number, which slightly increases $\eps'/\eps$,
has been already incorporated in some of the
predictions in Table~\ref{tab:pred}.

\section{CHIRAL PERTURBATION THEORY}
\label{sec:ChPT}

Below the resonance region
one can use global symmetry considerations to define another
EFT in terms of the QCD Goldstone bosons
($\pi$, $K$, $\eta$). The $\chi$PT formulation of the Standard Model
\cite{WE:79,GL:85,EC:95} describes
the pseudoscalar--octet dynamics, through a perturbative expansion 
in powers of momenta and quark masses
over the chiral symmetry breaking scale
($\Lambda_\chi\sim 1\; {\rm GeV}$).

Chiral symmetry fixes the allowed $\chi$PT operators.
At lowest order in the chiral expansion,
the most general effective bosonic Lagrangian
with the same $SU(3)_L\otimes SU(3)_R$ transformation properties
as the short--distance Lagrangian \eqn{eq:Leff} contains three terms,
transforming as $(8_L,1_R)$, $(27_L,1_R)$ and $(8_L,8_R)$, respectively.
Their corresponding chiral couplings are denoted by
$g_8$, $g_{27}$ and $g_{EW}$.

The tree--level $K\to\pi\pi$ amplitudes generated
by the lowest--order $\chi$PT Lagrangian do not contain any strong phase:
\beqn
\lefteqn{A_0 =
-{G_F\over \sqrt{2}} V_{ud}V^\ast_{us}\,\sqrt{2} f_\pi} &&
 \nonumber\\
&&  %\mbox{}\times
\left\{\left(g_8+{1\over 9}\, g_{27}\right) (M_K^2-M_\pi^2)
 -{2\over 3} f_\pi^2 e^2 g_{EW}\right\}  ,
\nonumber\\
\lefteqn{A_2 =
  -{G_F\over \sqrt{2}} V_{ud}V^\ast_{us}\, {2\over 9} f_\pi
\, \left\{5\, g_{27}\, (M_K^2-M_\pi^2)\, \right. }\nonumber\\
&&\left. - 3 f_\pi^2 e^2 g_{EW}\right\} \, .
\label{TREE}
\eeqn
From the measured $K\to\pi\pi$ rates one gets \cite{PGR:86}
$|g_8|\approx 5.1$ and $|g_{27}|\approx 0.29$. The $g_{EW}$ term
is the low--energy realization of the electroweak penguin operator.

The only remaining problem is the calculation of the chiral couplings
from the effective short--distance Lagrangian \eqn{eq:Leff},
which requires
to perform the matching between the two EFTs.
This can be easily done in the large--$N_C$ limit of QCD
\cite{HO:74,WI:79}, because
in this limit the four--quark operators factorize into currents
which have well--known chiral realizations:
\beqn
\label{eq:c2}
\lefteqn{g_8^\infty =  {3\over 5}\,C_2-{2\over 5}\,C_1+C_4
  -16\,L_5 \left({\langle\bar q q\rangle(\mu)\over f_\pi^3}\right)^2
  C_6 \, ,}&&
\nonumber\\
\lefteqn{g_{27}^\infty = {3\over 5}\,(C_1+C_2) \, , }&&
\nonumber\\
\lefteqn{g_{EW}^\infty =  -3\,
\left({\langle\bar q q\rangle(\mu)\over e\,f_\pi^3}
\right)^2\, C_8 \, .}&&
\eeqn

Together with eqs.~\eqn{TREE}, these results are equivalent to the
standard large--$N_C$ evaluations of the $B_i$ factors.
In particular, for $\eps'/\eps$ where only the imaginary part of
the $g_i$ couplings matter [i.e. Im($C_i$)] eqs.~\eqn{eq:c2}
amount to $B_8^{(3/2)}\approx B_6^{(1/2)}=1$. Therefore, up to minor
variations on some input parameters, the corresponding $\eps'/\eps$
prediction, obtained at lowest order in both the $1/N_C$ and
$\chi$PT expansions, reproduces the published results of the Munich
\cite{munich} and Rome \cite{rome} groups.

The large--$N_C$ limit is only applied to the matching between
the 3--flavour quark theory and $\chi$PT,
as indicated in Figure~\ref{fig:eff_th}.
The evolution from the electroweak
scale down to $\mu < m_c$ has to be done without any unnecessary expansion
in powers of $1/N_C$; otherwise, one would miss large corrections
of the form ${1\over N_C} \ln{(M/m)}$, with $M\gg m$ two widely
separated scales \cite{BBG87}.
Thus, the Wilson coefficients contain the full $\mu$ dependence.

The large--$N_C$ factorization of the four--quark operators $Q_i$
($i\not=6,8$) does not provide any scale dependence.
Since the anomalous
dimensions of these operators vanish when $N_C\to\infty$ \cite{BBG87},
a very important ingredient is lost in this limit \cite{PI:89}.
To achieve a reliable expansion in powers of $1/N_C$,
one needs to go to the next order where this physics is captured
\cite{PI:89,PR:91}. This is the reason why the study of the $\Delta I=1/2$
rule has proved to be so difficult. Fortunately, these operators
are numerically irrelevant in the $\eps'/\eps$ prediction.

The only anomalous dimensions which survive when $N_C\to\infty$
are precisely the ones corresponding to $Q_6$ and $Q_8$
\cite{BG:87,BBG87}. Moreover, the $Q_6$ and $Q_8$ matrix elements
are well approximated by this limit \cite{PR:91,JP:94}.
These operators  factorize into colour--singlet
scalar and pseudoscalar currents, which are $\mu$ dependent.
This generates the factors [$m_q = m_u = m_d$]
\be
\langle\bar q q\rangle(\mu) \, =\, - {f_\pi^2\, M_\pi^2\over 2 m_q(\mu)}
\, =\, - {f_\pi^2\, M_K^2\over (m_s + m_q)(\mu)}
\ee
in eqs.~\eqn{eq:c2}, which exactly cancel the $\mu$ dependence of
$C_{6,8}(\mu)$ at large $N_C$ \cite{BG:87,BBG87,PI:89,PR:91,JP:94,dR:89}.
It remains of course a dependence at next-to-leading order.

Therefore, while there are large $1/N_C$ corrections to Re($g_I$)
\cite{PR:91}, the large--$N_C$ limit is expected to give a
good estimate of Im($g_I$).

\section{CHIRAL LOOP CORRECTIONS}
\label{sec:loops}

The lowest--order calculation does not provide any strong phases
$\delta_I$. Those phases originate in the
final rescattering of the two pions and, therefore, are generated by
chiral loops which are of higher order in both the momentum
and $1/N_C$ expansions.
Analyticity and unitarity require the presence of a corresponding
dispersive FSI effect in the moduli of the isospin amplitudes.
Since the S--wave strong phases are quite large,
specially in the isospin--zero case,
one should expect large higher--order unitarity corrections.

The one--loop analyses of $K\to 2 \pi$ \cite{KA91,BPP,PA:00}
show in fact that pion
loop diagrams provide an important enhancement of the $\cA_0$ amplitude,
implying a sizeable reduction ($\sim 30\% $) of the fitted $|g_8|$
value. This chiral loop correction destroys the accidental numerical
cancellation in eq.~\eqn{EPSNUM}, generating a sizeable enhancement
of the $\eps'/\eps$ prediction \cite{PP:00a}.
However, the phase-shift $\delta_0$ predicted  at one--loop
is still lower than its measured value, which indicates that
a further enhancement should be expected at higher orders.

The large one--loop correction to $\cA_0$ has its origin in the
strong FSI of the two pions in S--wave, which
generate large infrared logarithms involving the light
pion mass \cite{PP:00b}. Identical logarithmic contributions appear
in the scalar  pion form factor \cite{GL:85},
where they completely dominate the $O(p^4)$ $\chi$PT correction.

Using analyticity and unitarity constraints \cite{GP:97},
these logarithms can be exponentiated to all orders in
the chiral expansion \cite{PP:00a,PP:00b}.
For the CP--conserving amplitudes, where the $e^2 g_{EW}$ corrections
can be safely neglected, the result can be written as
\beqn\label{eq:OMNES_WA}
\cA_I & = & \left(M_K^2-M_\pi^2\right) \; a_I(M_K^2)
\no\\
& = & \left(M_K^2-M_\pi^2\right) \; \Omega_I(M_K^2,s_0) \; a_I(s_0)\, ,
\eeqn
where $a_I(s)$ denote reduced off-shell amplitudes with
$s\equiv \left(p_{\pi_1}+p_{\pi_2}\right)^2$.
The Omn\`es exponential\footnote{
%%%%%%%
Equivalent expressions with an arbitrary number of subtractions for the
dispersive integral can be written \protect\cite{PP:00b}.
}
%%%%%%%
%
\beqn\label{eq:omega}
\lefteqn{\Omega_I(s,s_0) \;\equiv\;
e^{i\delta_I(s)}\; \Re_I(s,s_0)} &&
\\ && =
 \exp{\left\{ {(s-s_0)\over\pi}\int
{dz\over (z-s_0)} {\delta_I(z)\over (z-s-i\epsilon)}\right\}}
\no
\eeqn
provides an evolution of $a_I(s)$ from an arbitrary
low--energy point $s_0$ to $s=M_K^2$.
The physical amplitude $a_I(M_K^2)$ is of course independent of the
subtraction point $s_0$.

Taking the chiral prediction for $\delta_I(z)$ and expanding
$\Omega_I(M_K^2,s_0)$ to $O(M_K^2-s_0)$, one just reproduces the one--loop
$\chi$PT result. Eq.~\eqn{eq:omega} allows us to get a much more accurate
prediction, by taking $s_0$ low enough that the $\chi$PT corrections
to $a_I(s_0)$ are
small and exponentiating the large logarithms with the Omn\`es
factor. Moreover, using the experimental phase-shifts in the dispersive
integral one achieves an all--order resummation of FSI effects.
The numerical accuracy of this exponentiation has been successfully
tested \cite{PP:00b} through an analysis of the scalar pion form factor,
which has identical FSI than $\cA_0$.

\section{NUMERICAL PREDICTIONS}
\label{sec:numerics}

At $s_0 =0$, the chiral corrections are rather small.
To a very good approximation \cite{PPS:00}, we can just multiply
the tree--level $\chi$PT result for $a_I(0)$
with the experimentally determined Omn\`es exponentials \cite{PP:00b}:
\beqn
\Re_0(M_K^2,0) & = & 1.55 \pm 0.10\, ,
\no\\
\Re_2(M_K^2,0) & = & 0.92 \pm 0.03\, .
\eeqn

The corrected value of $\eps'/\eps$ is easily obtained from
eq.~\eqn{EPSNUM}, remembering that the lowest--order result was
equivalent to $B_8^{(3/2)}\approx B_6^{(1/2)}=1$. Therefore,
\beqn
\lefteqn{B_6^{(1/2)}  \, =\,  \left. B_6^{(1/2)}\right|_{N_C\to\infty}
\,\times\,\Re_0(M_K^2,0) \, =\, 1.55\, , }&&
\no\\
\lefteqn{B_8^{(3/2)}\, = \,\left. B_8^{(3/2)}\right|_{N_C\to\infty}
\,\times\,\Re_2(M_K^2,0)\, = \, 0.92\, , }&&
\no\\
\lefteqn{\Omega_{IB} \,\approx\,  0.16 \,\times \,
\Re_2(M_K^2,0)/\Re_0(M_K^2,0)
\, = \, 0.09 \, . } &&
\no
\eeqn
The isospin--breaking correction turns out to be small because the term
$B_6^{(1/2)}\,\Omega_{IB}$ corresponds to two final pions with $I=2$.
This is in good agreement with the value
$\Omega_{IB} = 0.08\pm 0.05$ recently obtained in ref.~\cite{MW:00}.

The large FSI correction to the $I=0$ amplitude gets reinforced
by the mild suppression of the $I=2$ contributions. The net effect
is a large enhancement of $\eps'/\eps$ by a factor 2.4,
pushing the predicted central value from $7\times 10^{-4}$
\cite{munich,rome} to $17\times 10^{-4}$ \cite{PP:00b}.
A more careful analysis, taking into account all hadronic and
quark--mixing inputs \cite{PPS:00} gives the
Standard Model prediction:
\be\label{eq:SMpred}
\varepsilon'/\varepsilon = (17\pm 6) \times 10^{-4}\, ,
\ee
which compares well with the present experimental world average
\cite{na48:00} in eq.~\eqn{eq:exp}.

\section{DISCUSSION}
\label{sec:summary}

The most important ingredients in
the Standard Model prediction of $\eps'/\eps$   %\eqn{eq:SMpred}
are indicated in Figure~\ref{fig:eff_th}. $\chi$PT fixes the
long--distance structure of the isospin $K\to\pi\pi$ amplitudes.
All short--distance information is contained in the local
chiral couplings.

The short--distance logarithms
are summed up with the OPE \cite{buras1,ciuc1}, while
%at long distances
the large infrared logarithms, involving the light pion mass,
are exponentiated through the Omn\`es factor and the measured
$\pi$-$\pi$ phase-shifts \cite{PP:00a,PP:00b}.

Once all important sources of large logarithms are kept under control,
the large--$N_C$ limit is used to match the short--distance
Lagrangian \eqn{eq:lag} with $\chi$PT and determine the chiral couplings
\cite{PPS:00}.
The local next-to-leading $\chi$PT contributions have been computed
for all operators $Q_i$ ($i\not= 6$). While $Q_8$ has small
corrections, the corresponding $O(p^4)$ correction to $Q_6$ is not
yet known since it involves couplings of the $O(p^6)$ strong Lagrangian.

The leading terms in $1/N_C$ approximate well the relevant operators
$Q_6$ and $Q_8$ \cite{PI:89,PR:91,JP:94}.
Therefore, it is reasonable
to expect that the size of missing higher--order $1/N_C$ corrections
will be around 30\%.
At present, the estimate of these next-to-leading $1/N_C$
contributions can only be done within specific models
\cite{trieste,dortmund,BP:00,PR:91,KPR:98}.

To summarize, using a well defined computational scheme,
it has been possible to pin down the value of $\eps'/\eps$
with an acceptable accuracy of about 40\%.
Within the present uncertainties, the resulting
Standard Model theoretical prediction \eqn{eq:SMpred}
is in good agreement with the measured experimental
value \eqn{eq:exp}, without any need to invocate a new
physics source of CP violation.

%%%%%%%%%%%%%%%%%%%%%%%%%%%%%%%
%\section*{ACKNOWLEDGEMENTS}
%\vspace{0.5cm}
\vfill

I warmly thank Elisabetta Pallante and Ignazio Scimemi
for a rewarding collaboration.
This work has been supported by the ECC, TMR Network
$EURODAPHNE$ (ERBFMX-CT98-0169), and by
DGESIC (Spain) under grant No. PB97-1261.

%%%%%%%%%%%%%%%%%%%%%%%%%%%%  REFERENCES %%%%%%%%%%%%%%%%%%%%

\end{document}